\documentclass[twoside]{article}
\usepackage{fleqn,espcrc2,epsfig}

\usepackage{graphicx}
\usepackage[figuresright]{rotating}

\newcommand{\AmS}{{\protect\the\textfont2
  A\kern-.1667em\lower.5ex\hbox{M}\kern-.125emS}}

\title{The unquenched $\Upsilon$ spectrum}

\author{L. Marcantonio\address[GLA]{Department of Physics and Astronomy,
        University of Glasgow, Glasgow, G12 8QQ, UK}, 
        P. Boyle\addressmark,
        C. T. H. Davies\addressmark,
        J. Hein\address{Newman Laboratory of Nuclear Studies, Cornell 
         University, Ithaca NY 14853, USA},
        and
        J. Shigemitsu\address[OSU]{Department of Physics, the Ohio State 
                University, OH 43210 USA}.
        UKQCD collaboration.}

\newcommand{\figwidth}{6.85cm}
\begin{document}

\begin{abstract}
We describe the bottomonium spectrum obtained on the UKQCD dynamical ensembles and its comparison 
to quenched results. We include a determination of $\alpha_s$ and $m_b$ from the 
dynamical results. 
\vspace{1pc}
\end{abstract}

\maketitle

\section{Introduction}
The $\Upsilon$ spectrum has long been recognised as a place
to search for effects from including dynamical quarks because
very precise calculations can be done and several mass splittings 
are sensitive to the short distance physics expected to 
be affected even by relatively heavy dynamical quarks. 
The calculations also enable the extraction of two important parameters of QCD, $\alpha_s$ and $m_b$.
There have been a number of calculations of the bottomonium spectrum using a range of dynamical configurations
~\cite{ouralpha,sesam,cppacs}. The results have not been conclusive, however.

Here we give new results using dynamical configurations from the UKQCD collaboration~\cite{irving}. These were obtained with Wilson (plaquette) glue and two flavours of clover dynamical quarks using a value of $c_{sw}$ given by the Alpha collaboration. Three ensembles are available with different dynamical quark masses. These ensembles also have the advantage over previous calculations (for investigations of this kind) of having the same lattice spacing. We believe that this enables systematic effects as a function of the physical dynamical quark mass to be isolated more easily and not confused with effects that arise as the lattice spacing changes when the quark mass is altered at fixed beta. The dynamical quark mass varies from around $m_s$ to around 2$m_s$ and roughly 100 independent configurations are available in each ensemble (separated by 40 trajectories).

We use the standard $\cal{O}$$(v_b^4)$ NRQCD action for the $b$ quarks~\cite{ourups}
but we also include discretisation corrections to the $\vec{E}$ and $\vec{B}$ fields through 
the use of $\tilde{F}_{\mu\nu}$. This is given (tadpole-improved) by
\begin{eqnarray}
\tilde{F}_{\mu\nu}(x) &=&\frac {5} {3} F_{\mu\nu}(x) \nonumber \\
&-& \frac{1} {6} [ \frac {1} {u_0^2} U_{\mu}(x)F_{\mu\nu}(x+a)U_{\mu}^{\dag}(x) \nonumber \\
&+& U^{\dag}_{\mu}(x-a)F_{\mu\nu}(x-a)U_{\mu}(x-a) \nonumber \\
&-& (\mu \leftrightarrow \nu)]
\end{eqnarray}
The use of improved fields should reduce discretisation errors in the fine structure on these relatively coarse lattices~\cite{ourscaling}. 
Two values of the bare quark mass $m_Qa$ were taken on each ensemble. They were chosen to give kinetic masses for the $\Upsilon$ particle which bracket the physical $\Upsilon$ mass, with the lattice spacing fixed from the splitting between the $1^1P_1$ and $1^3S_1$ states. This means that we can interpolate accurately to the physical $\Upsilon$ on each ensemble, again an important point in obtaining accurate results for the fine structure. 

We compare results on the dynamical ensembles to those on a reference quenched set at $\beta$ = 6.0, a slightly finer lattice spacing.

\section{Results for the spectrum}
Figure 1 plots the ratio of $a^{-1}$ obtained from the 1P-1S splitting in the $\Upsilon$ system to that defined from $r_0$ both on the dynamical configurations and the quenched set~\cite{irving}. We have previously checked that the ratio on quenched configurations is independent of the lattice spacing to within the statistical error. The results from the dynamical configurations are plotted against the square of the pseudoscalar light-light meson mass where the valence quark mass is equal to the sea quark mass. There is a possible sign that the discrepancy between the lattice spacing values obtained in the quenched approximation is reduced on the dynamical configurations. This is something that we would hope for, because in the real world there is only one value of the lattice spacing. 

\begin{figure}
\centerline{\epsfig{file=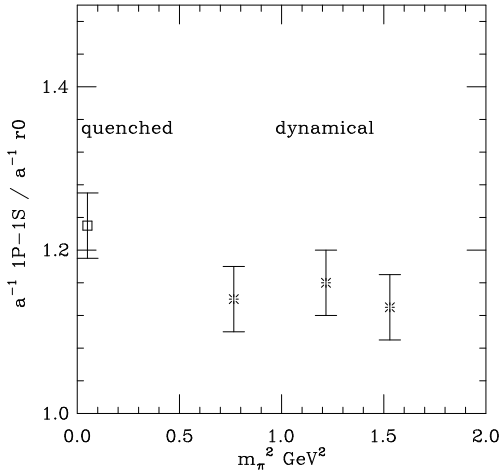,width=\figwidth}}
\caption{\label{arat} The ratio of the inverse lattice spacing obtained
from the 1P-1S splitting of the $\Upsilon$ to that obtained from 
$r_0$ on quenched and unquenched configurations. The results on 
unquenched configurations are plotted against the dynamical quark 
mass. }
\end{figure}

Figure 2 shows the hyperfine splitting between the $\Upsilon$ and the (so far unseen)
 $\eta_b$. In perturbation theory in a potential model this splitting is proportional to $|\psi(0)|^2$, 
and is therefore dominated by very short distances. 
We then expect it to increase on unquenching as the screening of $\alpha_s$ by dynamical quarks causes the potential at the origin to be deeper. This is clearly seen in our results, especially once the dynamical results are extrapolated to lighter values of the dynamical quark mass. 

\begin{figure}
\centerline{\epsfig{file=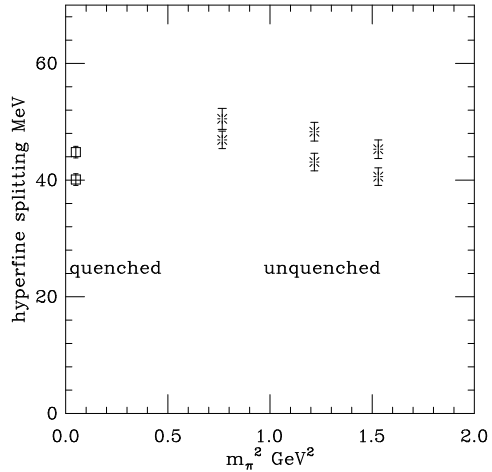,width=\figwidth}}
\caption{\label{hyps} The hyperfine splitting in MeV plotted 
for quenched configurations and (against dynamical quark mass)
for the dynamical configurations. Two values of heavy quark 
mass are given on each configuration and the physical 
$b$ quark lies between them. }
\end{figure}

The fine structure in the $P$ wave sector is much noisier and no clear effect of unquenching can be seen. This is evident in Figure 3.

\begin{figure}
\centerline{\epsfig{file=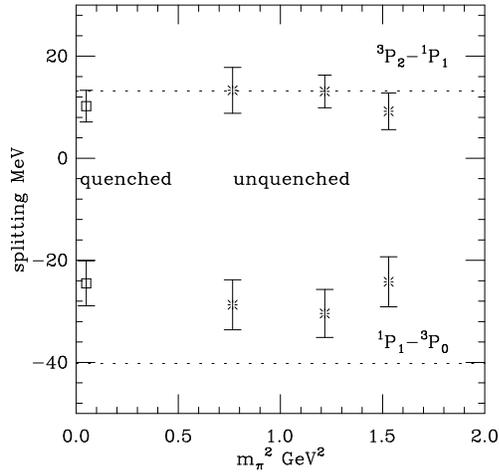,width=\figwidth}}
\caption{\label{psplits} Splittings among $\Upsilon$ P-wave states
on quenched and unquenched configurations. Dashed lines give 
experimental results. }
\end{figure}

\section{Results for $\alpha_s$ and $m_b$}

We use the standard method of determining $\alpha_s$ from a perturbative expansion 
of the plaquette~\cite{ouralpha,ournewalpha,sesamalpha}, setting the scale using the $\Upsilon$ 1P-1S splitting. 
This requires a calculation of the $N_f$ dependent part of the plaquette
at $\cal{O}$$(\alpha_s^2)$ for clover fermions. In the perturbative expansion to this order
$c_{sw}$ = 1.  We have (using also~\cite{weisz} and setting the dynamical quark mass to zero) 
\begin{eqnarray} 
-{\rm log}<\frac {1} {N_c} {\rm Tr} U_P > &=& \frac {4 \pi} {3} \alpha_P (1-b\alpha_P) \nonumber \\
\alpha_P &\equiv& \alpha_P(3.40/a)
\end{eqnarray} 
with
\begin{eqnarray}
b &=& b_1 N_c + b_2 N_f \\
b_1 &=& 0.39687 \\
b_2 &=& \frac {12 {\rm log} 3.40 - 10} {36 \pi} \\
&-& 4\pi P_4 + 32\pi x_2 \\
P_4 &=& 0.006696001 - 0.0050467 c_{sw} \\
&+& 0.029843 c_{sw}^2 \\
x_2 &=& 0.00069292 - 0.0000202 c_{sw} \\
&+& 0.00059624 c_{sw}^2 
\end{eqnarray}

Figure 4 shows our results with the $\alpha_s$ values run to a 
common scale of 8.2 GeV. Unquenching shows up very 
clearly as $\alpha^{(2)}$ is larger than 
$\alpha^{(0)}$. Comparison to other results shows a possible
dependence on the type of dynamical quarks which we are 
investigating further. 

\begin{figure}
\centerline{\epsfig{file=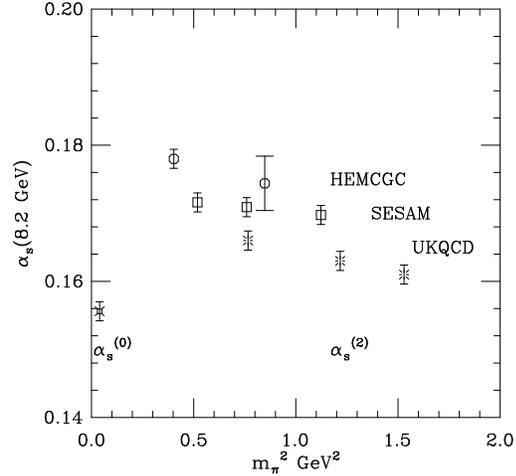,width=\figwidth}}
\caption{\label{alpha} The strong coupling constant determined
on various dynamical configurations at various dynamical 
masses compared to that obtained on quenched configurations.}
\end{figure}

The $b$ quark mass can be extracted using two different methods 
in NRQCD, either from the lattice bare quark mass or from 
the binding energy for, say, the $\Upsilon$. To convert 
to the mass in the $\overline{MS}$ scheme we need either 
the lattice mass renormalisation or the energy shift. Both 
of these are known only to $\cal{O}$$(\alpha_s)$~\cite{morningstar} at present
so our results here are not very precise. We see no effect 
from unquenching, nor do we see any variation with dynamical 
quark mass, see Figure 5. The most accurate results for $m_b$ come 
at present from the binding energy of the $B$ meson in 
the static limit~\cite{lubicz}.  

\section{Conclusions}

Effects of unquenching are seen on the $\Upsilon$ hyperfine splitting using the UKQCD dynamical configurations. The fact that they are matched in lattice spacing has helped with this. 

$\alpha_s$ can be extracted using the experimental $\Upsilon$ spectrum and a perturbative expansion for the plaquette. This will give a value for $\alpha_s$ to be used, for example, in perturbative matching of matrix elements to continuum QCD on the dynamical ensembles. 

\begin{figure}[t]
\centerline{\epsfig{file=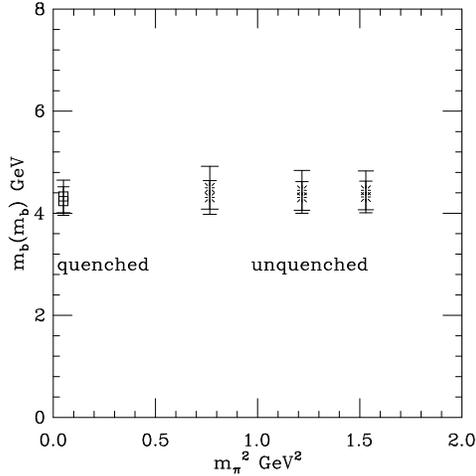,width=\figwidth}}
\caption{\label{mb} The $b$ quark mass in the 
$\overline{MS}$ scheme at its own scale, determined 
from the binding energy of the $\Upsilon$. 
Lattice and continuum perturbation theory to 
$\cal{O}$$(\alpha_s)$ are used. The results on the dynamical 
configurations are plotted against the dynamical quark 
mass. Two values are given on each ensemble corresponding to 
the bare lattice quark masses used.}
\end{figure}

\section{Acknowledgements}

We are grateful to the UK PPARC and the US DoE and NSF for funding for this work.

\end{document}